\documentclass[fleqn,usenatbib]{mnras}

\usepackage[T1]{fontenc}

\DeclareRobustCommand{\VAN}[3]{#2}
\let\VANthebibliography\thebibliography
\def\thebibliography{\DeclareRobustCommand{\VAN}[3]{##3}\VANthebibliography}

\usepackage{graphicx}	% Including figure files
\usepackage{amsmath}	% Advanced maths commands

\usepackage[T1]{fontenc}
\usepackage[utf8]{inputenc}

\usepackage{ae,aecompl}
\usepackage{graphicx}   % Including figure files
\usepackage{amsmath}    % Advanced maths commands
\usepackage{amssymb}    % Extra maths symbols

\usepackage{mathtools}
\usepackage{natbib}
\bibpunct{(}{)}{;}{a}{}{,} % to follow the A&A style

\usepackage{booktabs,caption}
\usepackage[flushleft]{threeparttable}

\hypersetup{
   colorlinks=true,
   linkcolor=blue,
   citecolor=blue,
   urlcolor=blue,
}

\usepackage{enumerate}

\usepackage{multirow}

\usepackage{array}
\newcolumntype{L}[1]{>{\raggedright\let\newline\\
\arraybackslash\hspace{0pt}}m{#1}}
\newcolumntype{C}[1]{>{\centering\let\newline\\
\arraybackslash\hspace{0pt}}m{#1}}
\newcolumntype{R}[1]{>{\raggedleft\let\newline\\
\arraybackslash\hspace{0pt}}m{#1}}
\usepackage{newtxtext,newtxmath}
\usepackage{hyperref}

\def\INSPIRE{\mbox{{\tt INSPIRE}}}

\def\sige{\mbox{$\sigma_{\rm e}$}}

\def\Msun{\mbox{M$_\odot$}}

\def\ML{\mbox{$M/L$}}

\def\mst{\mbox{$M_{\star}$}}

\def\lsim{\mathrel{\rlap{\lower3.5pt\hbox{\hskip0.5pt$\sim$}}
    \raise0.5pt\hbox{$<$}}}                % less than or approx. symbol
\def\gsim{~\rlap{$>$}{\lower 1.0ex\hbox{$\sim$}}}

\def\Fig{\mbox{Figure~}}

\def\Tab{\mbox{Table~}}

\def\Sec{\mbox{Section~}}

\def\App{\mbox{Appendix~}}

\def\UCMGs{\mbox{\textsc{UCMGs}}}

\def\Re{\mbox{$R_{\rm e}$}}

\def\Remaj{\mbox{$R_{\rm e, maj}$}}

\def\lephare{\mbox{\textsc{le phare}}}
\def\twodphot{\mbox{\textsc{2dphot}}}

\def\autoprof{\mbox{\textsc{autoprof}}}

\defcitealias{Tortora+18_UCMGs}{T18}
\defcitealias{Scognamiglio+20_UCMGs}{S20}

\title[KiDS~J0842+0059: 
the first fully confirmed relic beyond the local Universe] {INSPIRE: INvestigating Stellar Populations In RElics. IX. KiDS~J0842+0059: the first fully confirmed relic beyond the local Universe}

\author[C. Tortora et al.]
{C.~Tortora$^{1}$\thanks{E-mail: crescenzo.tortora@inaf.it}, 
G.~Tozzi$^{2}$, 
G.~Agapito$^{3}$, 
F.~La~Barbera$^{1}$,
C.~Spiniello$^{5,1}$\thanks{E-mail: chiara.spiniello@physics.ox.ac.uk}, 
R.~Li$^{4}$, 
G.~Carlà$^{3}$, 
G.~D'Ago$^{6}$, 
E.~Ghose$^{3}$, 
\newauthor
F.~Mannucci$^{3}$, 
N.~R.~Napolitano$^{7}$, 
E.~Pinna$^{3}$, 
M.~Arnaboldi$^{8}$, 
D.~Bevacqua$^{9,10}$, 
A.~Ferr\'e-Mateu$^{11,12}$, 
A.~Gallazzi$^{3}$, 
\newauthor
J.~Hartke$^{13,14}$, 
L.~K.~Hunt$^{3}$, 
M.~Maksymowicz-Maciata$^{17,5}$, 
C.~Pulsoni$^{15}$, 
P.~Saracco$^{9}$, 
D.~Scognamiglio$^{16}$, 
\newauthor
M.~Spavone$^{1}$ 
\\
$^{1}$ INAF -- Osservatorio Astronomico di Capodimonte, Salita Moiariello 16, 80131 - Napoli, Italy \\
$^{2}$ Max-Planck-Institut für Extraterrestrische Physik (MPE), Giessenbachstraße 1, D-85748 Garching, Germany \\
$^{3}$ INAF -- Osservatorio Astrofisico di Arcetri, Largo Enrico Fermi 5, I-50125 Firenze, Italy \\
$^{5}$ Sub-Dep. of Astrophysics, Dep. of Physics, University of Oxford, Denys Wilkinson Building, Keble Road, Oxford OX1 3RH, United Kingdom \\
$^{4}$ Institute for Astrophysics, School of Physics, Zhengzhou University, Zhengzhou, 450001, China \\
$^{6}$ Institute of Astronomy, University of Cambridge, Madingley Road, Cambridge CB3 0HA, United Kingdom \\
$^{7}$ Department of Physics E. Pancini, University Federico II, Via Cinthia 6, 80126 Naples, Italy \\
$^{8}$ European Southern Observatory, Karl-Schwarzschild-Stra\ss{}e 2, 85748, Garching, Germany \\
$^{9}$ INAF - Osservatorio Astronomico di Brera, via Brera 28, 20121 Milano, Italy \\
$^{10}$ DiSAT, Università degli Studi dell’Insubria, via Valleggio 11, I-22100 Como, Italy \\
$^{11}$ Instituto de Astrof\'isica de Canarias, Vía Láctea s/n, E-38205 La Laguna, Tenerife, Spain \\
$^{12}$ Departamento de Astrofísica, Universidad de La Laguna, E-38200, La Laguna, Tenerife, Spain \\
$^{13}$ Finnish Centre for Astronomy with ESO (FINCA), FI-20014 University of Turku, Finland \\
$^{14}$ Tuorla Observatory, Department of Physics and Astronomy, FI-20014 University of Turku, Finland \\
$^{15}$ Max-Planck-Institut f\"{u}r extraterrestrische Physik, Giessenbachstrasse, 85748 Garching, Germany \\
$^{16}$ Jet Propulsion Laboratory, California Institute of Technology, 4800, Oak Grove Drive - Pasadena, CA 91109, USA \\
$^{17}$ School of Physics, H.H. Wills Physics Laboratory, Tyndall Avenue, University of Bristol, Bristol, BS8 1TL, UK
}

\date{Accepted XXX. Received YYY; in original form ZZZ}

\pubyear{2025}

\begin{document}
\label{firstpage}
\pagerange{\pageref{firstpage}--\pageref{lastpage}}
\maketitle

\begin{abstract}
Relics are massive, compact and quiescent galaxies that assembled the majority of their stars in the early Universe and lived untouched until today, completely missing any subsequent size-growth caused by mergers and interactions. They provide the unique opportunity to put constraints on the first phase of mass assembly in the Universe with the ease of being nearby. While only a few relics have been found in the local Universe, the \INSPIRE\ project has confirmed 38 relics at higher redshifts ($z \sim 0.2-0.4$), fully characterising their integrated kinematics and stellar populations. However, given the very small sizes of these objects and the limitations imposed by the atmosphere, structural parameters inferred from ground-based optical imaging are possibly affected by systematic effects that are difficult to quantify. In this paper, we present the first high-resolution image obtained with Adaptive Optics Ks-band observations on SOUL-LUCI@LBT of one of the most extreme \INSPIRE\ relics, KiDS~J0842+0059 at $z \sim 0.3$. We confirm the disky morphology of this galaxy (axis ratio of $0.24$) and its compact nature (circularized effective radius of $\sim 1$ kpc) by modelling its 2D surface brightness profile with a PSF-convolved S\'ersic model. 
We demonstrate that the surface mass density profile of KiDS~J0842+0059 closely resembles that of the most extreme local relic, NGC~1277, as well as of high-redshift red nuggets. We unambiguously conclude that this object is a remnant of a high-redshift compact and massive galaxy, which assembled all of its mass at $z>2$, and completely missed the merger phase of the galaxy
evolution.
\end{abstract}


%%%%%%%%%%%%%%%%%%%%%%%%%%%%%%%%%%%%%%%%%%%%%%%%%%

%%%%%%%%%%%%%%%%% BODY OF PAPER %%%%%%%%%%%%%%%%%%

\section{Introduction}\label{sec:intro}

Relic galaxies are the final stage of high redshift passive,
massive and compact galaxies, the red nuggets \citep{Trujillo+09_superdense, Damjanov+09_rednuggets}, that evolved
undisturbed until the nearby Universe,
missing the second phase of the so called
``two-phase" scenario (\citealt{Oser+10}). According to this formation scenario, red nuggets form at high redshift through an intense and rapid series of dissipative processes. These processes produce a massive, passive, and very compact galaxy after star formation (SF) quenches \citep{Labbe+23}. Once this dissipative phase concludes (around $z\sim2$, \citealt{Zolotov+15_nuggets}), a second phase begins, characterized by gas inflows, dry mergers, and interactions. This accretion phase, which is significantly more extended in time than the first one, drives the dramatic structural evolution and size growth of early-type galaxies (ETGs) from 
$z\sim2$ to the present day (\citealt{Naab+09, Buitrago+18_compacts}).
However, since mergers are stochastic events, a small percentage (1–10 per cent) of high-redshift red nuggets could miss completely the size-growth phase, and evolve undisturbed and passively in cosmic time \citep{Hopkins+09_compacts,Quilis_Trujillo13}. These nearby objects are the ancient relics of the early Universe.

In the local Universe,
only few relic galaxies are known and have been studied in detail
(e.g., \citealt{Trujillo+14}; \citealt{Ferre-Mateu+15,
Ferre-Mateu+17, Yildrim17, Beasley18, Martin-Navarro19, Salvador-Rusinol22,Comeron23}), pointing to an abundance consistent with
theoretical expectations (e.g., \citealt{Quilis_Trujillo13,Wellons16,Flores-Freitas22,Moura+24}). Since 
these galaxies are expected to be slightly  more common at higher redshifts,
efforts have been dedicated to explore redshifts up to 
$z \sim 0.7$, {using} HST data (e.g., \citealt{Damjanov+15_compacts}), or wide-field ground-based images (e.g.,
\citealt{Tortora+16_compacts_KiDS,Tortora+18_UCMGs,Tortora+20_UCMGs_env};
\citealt{Charbonnier+17_compact_galaxies};
\citealt{Buitrago+18_compacts}; \citealt{Scognamiglio+20_UCMGs}; \citealt{Lisiecki+23}). This has allowed the creation of a statistically large sample of relic candidates, i.e., ultra-compact massive galaxies (UCMGs, effective radius $\Re \lsim 2 \, \rm kpc$ and stellar mass $\mst \gsim \,  6 \times 10^{10} \, \Msun$) characterised by very red colors.

With the INvestigating Stellar Populations In RElics (\INSPIRE) Project\footnote{Based on the ESO Large Program ID: 1104.B-0370, PI: C. Spiniello.}, we obtained the
first systematic confirmation of relic galaxies beyond the local Universe. Thanks to high signal-to-noise, UVB-to-NIR X-Shooter (XSH) spectra, we obtained precise determinations of the stellar age, metallicity, [Mg/Fe], Initial Mass Function (IMF) and velocity dispersion for the full sample of 52 UCMGs (\citealt{Spiniello+20_INSPIRE,Spiniello+21_INSPIRE-DR1,Spiniello+21_INSPIRE_Messenger,Spiniello+23_INSPIRE-V, DAgo+23_INSPIRE-III, Martn-Navarro+23_ISPIRE-IV, Maksymowicz-Maciata+24_INSPIRE-VI,Scognamiglio+24_INSPIREVII}). Out of these, 38 assembled more than 75\% of their mass by $z=2$, and hence have been classified as relic galaxies. This has more than tripled the number of currently known relic galaxies, also pushing the redshift boundaries, and confirming the existence of a ``degree of relicness'' (DoR; \citealt{Ferre-Mateu+17}) in terms of the star formation history (SFH).
Indeed, from a detailed stellar population study of the final \INSPIRE\ sample, in \citet{Spiniello+23_INSPIRE-V} we have quantified the DoR as a dimentionless number, varying from 1 to 0, and  defined in terms of a) the fraction of stellar mass formed by $z = 2$, b) the time at which a galaxy has assembled 75 per cent of its mass, and b) the final assembly time. A high DoR ($>0.7$) indicates a very early complete mass assembly, with almost 
no contribution from later star formation episodes. A lower DoR ($\leq0.34$) means that there is a non-negligible percentage of populations with younger ages and different metallicities, and hence a much later time of final assembly.

\INSPIRE\ has been the first important step to
understand the physical processes which drove red nuggets formation at high redshift and their undisturbed evolution from $z\gsim \, 2$ until the local Universe, providing constraints on the in-situ stellar populations in the central regions of the most massive and passive galaxies. 
However, both the imaging used to select the galaxies\footnote{Multi-band images from the Kilo Degree Survey  \citep{deJong+15_KiDS_paperI,deJong+17_KiDS_DR3}.} and XSH spectroscopy 
to confirm these systems as relics are seeing-dominated. Given the very small sizes of these objects on sky, compared to the ground-based FHWM seeing (i.e., effective radii $\sim 0.2-0.5 \, \rm arcsec$ vs $\rm FWHM \sim 0.7 \, \rm arcsec$ in the best cases), the only viable way to securely estimate structural properties, constrain their morphology with high precision and confidence, and confirm the pureness of our selected sample is the use of Adaptive Optics (AO) supported imaging or space imaging data ($\rm FWHM \sim 0.1 \, \rm arcsec$). 

In this paper, the 9th
of the \INSPIRE\ series, we present the first AO observations, obtained with the 
Single conjugated adaptive Optics Upgrade for LBT (SOUL) imager \citep{Pinna+2019_SOUL} on the Large Binocular Telescope (LBT, \citealt{Hill10_LBT}), of one of the most extreme relic galaxies in our sample: KiDS~J0842+0059. At the time of the proposal submission, KiDS~J0842+0059 was the most promising among the \INSPIRE\ galaxies that could be followed up with LBT. From KiDS seeing-limited data, we computed an effective radius of $\Re = 1 \, \rm kpc$ averaging the values obtained from $g$, $r$ and $i$ images \citep{Scognamiglio+20_UCMGs}. Here, using the new AO observations, we unambiguosly confirm the compact nature of KiDS~J0842+0059, by determining more accurate size and structural parameters. This is the first time that a confirmed relic galaxy at these redshifts is observed with such high data quality and a spatial resolution of $\sim 0.1$ arcsec. Our study is the first step to confirm the truly compact nature of the \INSPIRE\ sample.

The paper is organized as follows. In \Sec\ref{sec:data} we present the  observations and the data analysis. Structural parameters are presented in \Sec\ref{sec:results}, alongside with the target and its characteristics, which confirm the compact nature of the target. In \Sec\ref{sec:profiles} we compute mass density profile of the target and compare it with these of other ultra-compact massive galaxies. A summary of the results is outlined in \Sec\ref{sec:conclusions}.

\section{Observations, data reduction and PSF modelling}\label{sec:data}

\begin{figure*}
\centering
\includegraphics[width=0.99\textwidth]{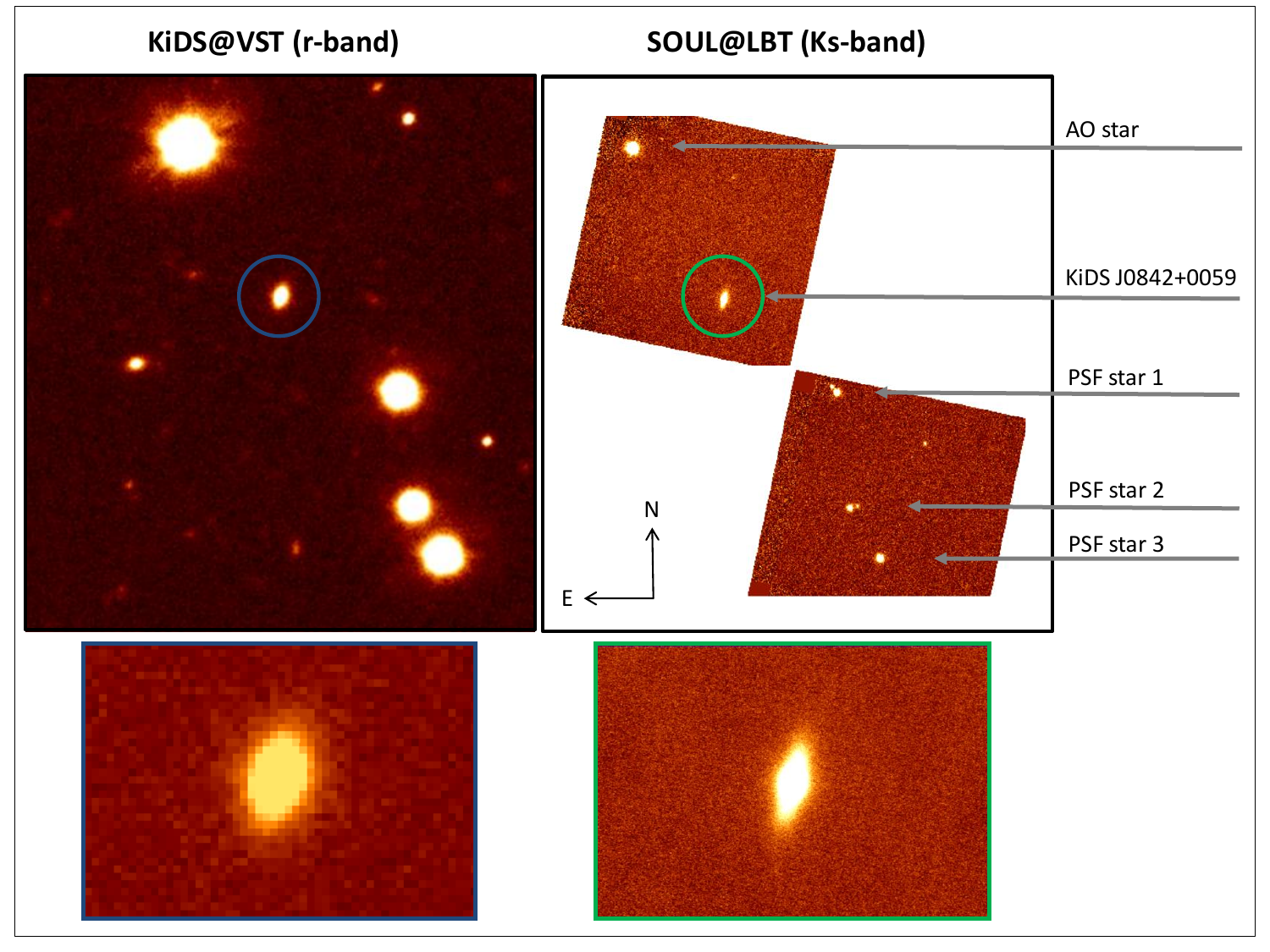}
\caption{\textit{Top.} Field of view of $\sim 1.1 \times 1.2 \, \rm arcmin^{2}$ including the target {(blue/green circle)}, the AO star and the three
stars used to determine the PSF of the SOUL data. 
We show the KiDS@VST r-band image
on the left panels and 
a SOUL Ks-band mosaic built merging the pointing on the AO star and the target and that containing the PSF stars on the right. Each SOUL pointing has a size of  $\sim 30 \times 30 \, \rm arcsec^{2}$. Both images are oriented with East and North along the horizontal and vertical directions. \textit{Bottom.} Cutouts of $\sim 8\times 11\, \rm arcsec^{2}$ from KiDS (left) and SOUL (right) centered on KiDS~J0842+0059.}
\label{fig:field}
\end{figure*}

The images analysed in this paper have been obtained on January 2021 with the SOUL imager that feeds the instrument LUCI (\citealt{Heidt+18_LUCI}) on LBT,
in the Ks filter. These observations are part of the proposal 
INVEstigating Relic galaxies with SOUL (INVERSO, P.I.: C. Tortora). The pixel scale of the imager is of $0.015 \, \rm arcsec/pix$.

\subsection{Observations}

The single-conjugate (SCAO) systems mounted on LBT use a natural guide star (NGS) to drive the deformable mirror. Under median seeing conditions, Strehl ratios (SR) above 0.1 can be obtained in the Ks band for NGS (AO star in the following) brighter than $R=15$ mag and separations below 40 arcsec, limiting the number of observable targets \citep{Pinna+23_SOUL}. Furthermore, bright closer stars are necessary to carefully map the point-spread function (PSF) of the instrument. The further away the target is from the AO star, the worse is the modeling of the PSF and the larger is the distortion of the image along the direction of the star. Hence, the ideal strategy when using AO systems is to use PSF stars at a distance from the AO star comparable to that of the target.

Only four objects from the current \INSPIRE\ sample have a star in their proximity, bright enough to be used as AO star. At the time of the proposal submission, KiDS~J0842+0059 was the most interesting of the four targets with a star in their proximity. Indeed, this is one of the most extreme relics of the sample, having formed more than 98\% of its stellar mass within $\sim3$ Gyr after the Big Bang, and has 3 nearby bright stars to be used for PSF modelling. However, unfortunately these stars are further away from the AO star than the relic galaxy. 
In particular, the AO star is located at a distance of 23.2 arcsec from the target. It has a Gaia-band magnitude of 15.1 mag (KiDS $g$- and $r$-band of 15.5 mag and 15.1 mag, respectively, and 2MASS K-band of 13.8 mag). The three PSF stars are instead at the distances of 42.3 ($K=15$ mag), 55.4 ($K=15.4$ mag) and 62.9 ($K=14.6$ mag) arcsec from the AO star.

The LUCI observations consist of two data sets relative to two different sky pointings: one centred on the target galaxy and the AO star, and the other one containing the other stars used to estimate the PSF, as described below. In each field, the main target (i.e., either the galaxy or the stars) was dithered on the field of view, hence observed in different positions on the LUCI detector\footnote{The dithering has been performed with shifts in the range 2-3.5 arcsec from the reference position of the target along two orthogonal axes.}. For each dithered target position, an observing shot (or frame) of $NDIT=12$ exposures of 10 s each was carried out. We observed 27 frames for the pointing centered on the target and the AO star, and 2 frames for the pointing containing the three PSF stars.

\subsection{Data reduction}

We reduced and combined the data frames of the galaxy and, separately, of the PSF stars, by means of our custom-made python routine ({\tt PySNAP}), which extensively makes use of functions from the $\mathtt{ccdproc}$ package\footnote{All functions mentioned in the text belong to this package.} included in the
$\mathtt{astropy}$ library. Our procedure performs sky
subtraction, accounts for the presence of bad pixels, and applies dark and flat field corrections. Being all on-target exposures, we produced a master sky frame using the $\mathtt{combine}$ function: all dithered on-target frames were combined by calculating the
median value pixel-by-pixel, and rejecting the most deviating pixels (those beyond $2 \sigma$) with a sigma clipping technique. A flat field image was created with the same combining method as for the sky frame. We then corrected for dark using a dedicated dark exposure and finally divided the combined image for the median values of its pixels. Finally, we used the resulting flat field frame as input of the $\mathtt{ccdmask}$ function to generate the bad pixel mask.

With these ingredients, for each 12-exposure shot we generated the mean frame, in order to increase the signal-to-noise ratio (S/N). All corrections were then applied to each mean frame, namely the master-sky subtraction and the corrections for dark, flat field and bad pixels. 
At this point, the different dithered, corrected mean frames were aligned first using the WCS offsets stored in the header file, then applying an additional shift correction by fitting a 2D elliptical Moffat profile to the target
(i.e., either the galaxy or the stars) SB. Although better suitable to reproduce the SB of point-like sources (e.g., stars) and not of extended sources, such as our galaxy, a 2D Moffat profile is however successful in identifying the galaxy emission centroid, necessary to align the distinct mean frames. Out of the 27 frame sequences where the galaxy appears, we discarded 3 sequences of clearly much worse quality, likely obtained in either bad atmospheric conditions or during instrumental issues (e.g., open loop of the AO system). The poor quality of these frames is confirmed by the larger FWHM (i.e., $\gsim 10$ pixels) of the 2D Moffat profile fitted to the AO star, present in all these frames, compared to the good galaxy frames where the AO star is also covered (FWHM $\sim 5$ pixels)\footnote{
Unfortunately, the AO star was not visible in all galaxy frames. As a result, we could neither use it to align the different frame sequences nor to thoroughly assess the quality of each sequence.}. For the galaxy, three frames were
rejected, for the pointing containing the three PSF stars all frames were used. 
Finally, as last step, we combined all non rejected frames for each pointing. The final combined images for each pointing were created as an average frame of the
corrected, aligned frames, using the $\mathtt{combine}$
function. In details, for the pointing including the target, we observed a total of 27 frames, and we combined 24 of these with optimal FWHM, totalling an
exposure time of 2880s. For the pointing including the PSF stars, we combined 2 frames of $12 \times 10$s each, hence reaching a  total exposure time of 240s. 

Along with the final combined image, we produced a weight map reporting pixel-by-pixel the value of $1/\sigma^2$, where $\sigma$ is the uncertainty on photon counts in a given pixel. Since in the near-IR the dominant uncertainty is due to the Poissonian noise, the weight map was created in the following way. We first associated a single noise value $\sigma_i$ to each corrected frame $i$, calculated as Median Absolute Distance (MAD) value among those pixels not contaminated by source emission. Then, the value of $\sigma$ in a given pixel $(x, y)$ of the final combined
image was computed as:
\begin{equation}
\sigma(x,~y)=\sqrt{\frac{\sum_{i=1}^{N}
\sigma_i^{2}(x,~y)}{N(x,~y)}},
\end{equation}
where N is the total number of combined frames in the pixel
$(x,~y)$, considering both the overlapping of different dithered
frames and masked bad pixels.

The KiDS@VST and SOUL@LBT field of views, oriented with East and North along the horizontal and vertical directions, and including target, AO star and PSF stars, are shown in \Fig\ref{fig:field}. The figure highlights the much higher resolution we achieved with AO-supported images.

\subsection{PSF modelling}\label{subsec:PSF}

Since the PSF FWHM varies with distance from the target, we conducted a detailed PSF modeling to accurately determine the PSF FWHM at the target’s location and assess whether the planned objectives in terms of FWHM have been met. This is crucial to assess the possible uncertainties and systematics that could affect the determination of the structural parameters of the galaxy, which is the main goal of this paper.

Our approach consisted of two steps. First, we fitted the 4 stars (AO and PSF stars) using \textit{TIPTOP} (\citealt{Neichel+21_TIPTOP}), and then we extrapolated the PSF to the galaxy position. Technical details are provided in App. \ref{app:PSF_full}.  

The circularized FWHM of the three PSFs is of $\sim 0.20$, $\sim 0.24$ and $\sim 0.25$ arcsec,
calculated using, independently, TIPTOP or (equivalently) fitting a Moffat profile to the PSF star 1, 2 and, 3, respectively. The extrapolated FWHM value to the target position is of $\sim 0.14$ arcsec, with a major axes value of 0.168 arcsec and  minor-axes of 0.116 arcsec. This is in line with the target value planned during the proposal preparation using the SOUL SR calculator\footnote{http://adopt.arcetri.inaf.it/strehl.html}.

\begin{figure*}
\centering
\includegraphics[width=0.99\textwidth]{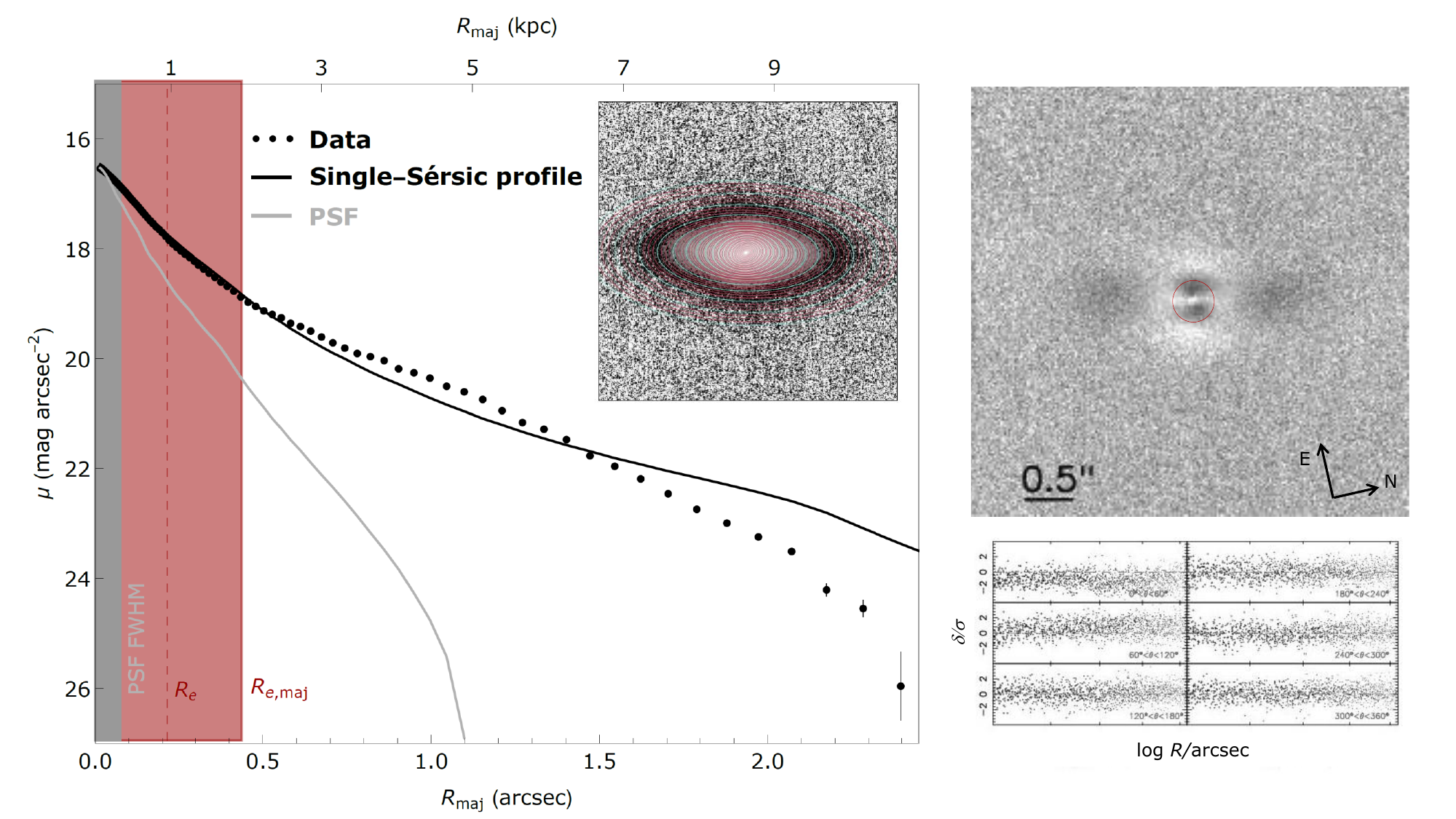}
\caption{{\it Left}. The azimuthally averaged (1D) SB profile, derived from the isophotes shown in the inset panel,  plotted as a function of the major axis \Re\ (black symbols with uncertainties). The solid black curve represents the best-fitting PSF-convolved model, obtained by fitting a 2D single-S\'ersic profile to the image using \twodphot. The parameters of this fit are listed in Table~\ref{tab:results}. The light red region marks the major-axis \Re, while the red vertical dashed line indicates the circularized \Re\ for the best-fitting single-S\'ersic profile. The gray curve corresponds to the PSF used in the fitting, and the gray shaded area represents the PSF's FWHM. The galaxy image in the inset was created with \autoprof, with isophotal ellipses overlaid. {\it Right}. Residual image generated by \twodphot\ for the single-S\'ersic fit. Darker (lighter) gray is for larger (smaller) residuals. The red circle marks the circularized \Re. The image is aligned according to the SOUL image, with the AO star located in the upper-right direction. The residual image is computed as the difference between the observed image and the best-fitting model, divided by the error in the pixel. In the six lower panels, the residuals of the galaxy flux per pixel, after model subtraction, are plotted as a function of distance from the galaxy center, separated into different bins of polar angle.}\label{fig:SBprofiles} 
\end{figure*}

\section{KiDS~J0842+0059: a fully confirmed relic beyond the local Universe}
\label{sec:results}

In this section, we first provide a detailed discussion of the characteristics of KiDS~J0842+0059, derived from past seeing-limited observations from KiDS and \INSPIRE. We then complement this analysis with the structural parameter analysis based on this new observations from LUCI-SOUL@LBT, which confirm its ultra-compact nature.

\subsection{Physical and stellar population properties}

KiDS~J0842+0059 is one of the most extreme relic galaxies in the \INSPIRE\ sample. This galaxy, 
at $z = 0.2959$, has formed $\sim 75 \%$ of its stellar mass within the first Gyr after the Big Bang 
and fully completed its assembly 3.5 Gyr after it. Within the \INSPIRE\ project, galaxies have been operationally classified in terms of their DoR into three broad families: non-relics (DoR$<0.34$, 14 objects), relics ($0.34\le\text{DoR}\le0.7$, 29 objects) and extreme relics (DoR$>0.7$, 9 objects). The DoR ranges between 0.06 and 0.83 with a median of 0.4. KiDS~J0842+0059 has a DoR of $\rm 0.73$, ranking as the seventh most extreme relic galaxy in the \INSPIRE\ sample.

KiDS~J0842+0059 has a stellar mass\footnote{The error on the stellar mass is of the order of $0.1$ dex.} of $\mst \sim 10^{11}\, \Msun$, inferred from Spectral Energy Distribution (SED) fitting in the KiDS $ugri$ bands using \lephare\ \citep[see more details in][]{Tortora+18_UCMGs,Scognamiglio+20_UCMGs}.
In \citet{Spiniello+23_INSPIRE-V}, we infer an integrated stellar velocity dispersion of $\sige \sim 324\pm 32 \rm \,km s^{-1}$, from a XSH spectra extracted over an aperture encapsulating 50\% of the light (R50)\footnote{This is likely a lower-limit for $\sige$, since the light contained in the R50 aperture is a mixture from inside and outside the effective radius, see App.~A in  \citet{Spiniello+21_INSPIRE-DR1} for  details.}. Like for the other extreme relics, the $\sige$ is larger than the average values computed for normal-sized galaxies and non-relics with similar stellar masses (see \Fig\ 9 in \citealt{Spiniello+23_INSPIRE-V}). 

The stellar population properties of KiDS~J0842+0059 fully resemble those of local relics: it has a very old mass-weighted age of $\sim 10 \, \rm Gyr$, compared to the age of the Universe at $z=0.3$ \citep{Spiniello+23_INSPIRE-V}, supersolar metallicity ([M/H] $= 0.36^{+0.03}_{-0.04}$) and [Mg/Fe] ($0.26^{+0.06}_{-0.06}$), and a dwarf-rich IMF with a low-mass end slope of $\Gamma_b=2.80^{+0.26}_{-0.32}$ \citep{Maksymowicz-Maciata+24_INSPIRE-VI}. 
When instead assuming a universal, Kroupa-like IMF (as in \citealt{Spiniello+23_INSPIRE-V}), the galaxy resulted equally old ($\sim 10 \, \rm Gyr$), even more Mg-enhanced ([Mg/Fe]$\sim0.4$) and still metal rich, although less super solar than when changing the IMF slope ([M/H] $\sim0.25$). These stellar population estimates allow us to update the stellar mass of KiDS~J0842+0059 obtained from photometry. Specifically, adopting the best-fit model from \cite{Spiniello+23_INSPIRE-V}, we find a mass-to-light ratio of $M/L_{\rm K} = 0.94$, which, combined with a total K-band magnitude of 17.6, yields a stellar mass of $\mst \sim 2.5 \times 10^{11}\, \rm \Msun$. Using instead the stellar population results from \cite{Maksymowicz-Maciata+24_INSPIRE-VI} with a variable IMF, leads to a doubled mass estimate of $\mst \sim 5 \times 10^{11}\, \rm \Msun$. However, here,  we will stick with our previous mass estimate from photometry, for uniformity with the rest of galaxies analyzed in the \INSPIRE\ sample. In a future publication, we will revise all the masses for the \INSPIRE\ sample. Finally, in terms of mass density profile, KiDS~J0842+0059 matches very well the local relic prototype NGC~1277 \citep{Ferre-Mateu+17} and high-z red nuggets \citep{Szomoru+12}.

\begin{table*}
\centering
\caption{KiDS~J0842+0059 structural parameters computed using KiDS and AO data for a single-S\'ersic profile. Angles are counted from west to north.}\label{tab:results}
\begin{tabular}{cccccccc}
\hline
\rm Data &  PSF & n & q & $\theta_{q}$ & $R_{e,maj}$ & $R_{e,circ}$ & $\chi^{2}$\\
\rm  & &   &  & (degree) & (arcsec) & (arcsec) & \\
\hline
KiDS &  2-Moffat &  3.27 & 0.29 & 76.8 & 0.420 & 0.226 & 1 \\ 
\hline
AO &  AO-Extrapolated  & 5.72 $\pm$ 0.5 & 0.24 $\pm$ 0.10 & 74 $\pm$ 2 & 0.437 $\pm$ 0.1 & 0.214 $\pm$ 0.001 & 1.4 \\ 
\hline
\end{tabular}
\end{table*}

Modelling ground-based KiDS optical images, we estimated an 
effective radius of 
$\Re = 1 \, \rm kpc$ \citep{Scognamiglio+20_UCMGs}, and therefore classified KiDS~J0842+0059 as ultra-compact. However, as shown in Appendix B of \cite{Tortora+18_UCMGs}, while KiDS effective radii are statistically precise and accurate on a population-level, individual cases — particularly those with the most compact sizes — exhibit larger uncertainties (up to $\sim50\%$), and results can be affected by potential biases. Therefore, the main scope of this paper is to get an accurate measurement of the structural parameters of KiDS exploiting the accuracy of the unprecedented AO observations described in \Sec\ref{sec:data}. A much higher precision is indeed achievable thanks to a  PSF which is around 6 times narrower than that of ground-based, seeing-limited data, and hence comparable to the effective radius of the galaxy. The following subsection presents the main results of the paper: a precise estimate of the structural parameters (S\'ersic index, axis ratio and effective radius) from the AO-supported NIR images.

\subsection{Structural parameter estimates from AO observations}

We first performed a preliminary analysis, to characterize in a non-parametric way the shape of the galaxy, by running the code \autoprof\  \citep{Stone+2021_Autoprof} on the AO-supported image. \autoprof\ fits elliptical isophotes to the image and then performs a non-parametric extraction of the azimutally averaged light profile of the galaxy. The image of the galaxy is shown in an inset on the left panel of \Fig\ref{fig:SBprofiles}, together with the fitted isophotes; for visualization purposes, the isophotes are colored in red, with one in cyan every fourth isophote. These are rounder in the center ($q=0.8$) and more elongated in the outskirts ($q=0.45$). The artificial twist in the position angle of the ellipses is also clear, with the central ($\lsim 0.2''$) isophotes, where the effect of the PSF is stronger, rounder and stretched towards the AO star (along the north-east direction). The AO higher-resolution image clearly  
shows the disky shape of the galaxy isophotes, which were blurred by the worse PSF in the KiDS image.

We used \twodphot\ \citep{LaBarbera_08_2DPHOT} to fit the target image with a PSF-convolved 2D S\'ersic model. The PSF used to convolve the S\'ersic profile, has been modelled as described in \Sec\ref{subsec:PSF} and discussed in greater detail in \App\ref{app:PSF_full}. We have also explored a more complex model and its impact on the fit using the combination of a S\'ersic and an exponential profile in \App\ref{app:double_sersic}. We have estimated 
the following parameters with their uncertainties: S\'ersic index ($n$), axis ratio ($q$), position angle ($\theta_{q}$), major axis and circularized effective radius ($R_{\rm e,maj}$ and $R_{\rm e,circ} = \sqrt{q} R_{\rm e,maj}$). The results of the fitting are compared with those obtained from KiDS seeing limited data and
summarized in \Tab\ref{tab:results}.

We have run \autoprof\ not only on
the galaxy image, but also on 
the image of the best-fitting models convolved with the PSF\footnote{The latter image is provided in output by \twodphot.}. The radial SB profiles for image (black symbols) and best-fitting models (blue lines) are shown in the left panel of \Fig\ref{fig:SBprofiles}.

By fitting a single-S\'ersic PSF-convolved profile we have found
$n = 5.7 \pm 0.5$,
$q = 0.24 \pm 0.1$ and, $R_{\rm e,circ} = 0.214 \pm 0.001$ arcsec, which corresponds to $\Re = 1 \, \rm kpc$.
The reduced $\chi^{2}$ is 1.4. All parameters are determined with high accuracy, except for the axis ratio, which exhibits the largest relative error of approximately $\sim 40\%$.

The residual image is shown in the right panel of \Fig \ref{fig:SBprofiles}. We find an asymmetrical residual concentrated in the very central galactic region ($\lsim 0.3$ arcsec), which can be probably connected to the inability to model the PSF with the needed accuracy. 
For example, imperfect PSF modeling, such as errors in the PSF slope, can influence the slope of the best-fitting profile and consequently the value of n. Nevertheless, the residuals calculated at different polar angles as a function of the distance from the center give a clear indication of a quite good fit. 

By calculating the growth curve from the data shown in \Fig\ref{fig:SBprofiles} and determining \Remaj\ as the radius enclosing half of the total light, we find $\Remaj = 0.45$ arcseconds\footnote{Since no PSF deconvolution has been applied, this value represents an upper limit.}. It is closely matching the estimate obtained from fitting the 2D S\'ersic profile to the image.

In \App\ref{app:double_sersic} we also explore a double-component fit. Although 
the central component seems to address our challenge in accurately modelling the PSF at the galaxy position and therefore a slightly lower $\chi^2$, we express reservations about the two-component model's ability to represent a physically meaningful solution for the relic. This consideration is justified by the fact that the central component specifically fits the central SB profile, where the PSF plays a crucial role. Additionally, such best-fitting model's axis ratio and orientation mirror the characteristics of the more central isophotes.

In conclusion, the circularized effective radius is robustly determined with a value of $\Re = 0.214 \pm 0.001$, which corresponds to $\Re \sim 1 \, \rm kpc$, confirming the KiDS seeing-dominated results reported in \cite{Scognamiglio+20_UCMGs}. We have also confirmed that the galaxy presents disky isophotes. Therefore, we can now securely conclude that J0842+0059 is an ultra-compact (massive) galaxy, with size and shape resembling the local relic prototype NGC~1277 and high-z red nuggets.

\begin{figure*}
\centering
\includegraphics[width=0.99\textwidth]{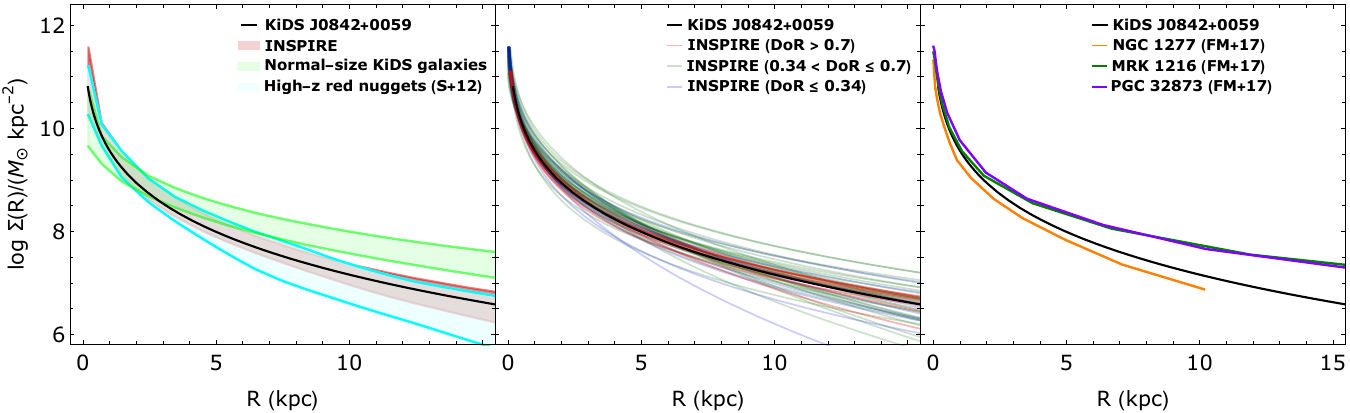}
\caption{Projected mass density profile of KiDS~J0842+0059 (black line). In the left panel, this is compared to the range around the median including 68\% of profiles for \INSPIRE\ galaxies \citep{Spiniello+23_INSPIRE-V}, normal-size KiDS galaxies \citep{Roy+18,Tortora+20_UCMGs_env} and high-z red-nuggets \citep{Szomoru+12}. In the middle panel, the mass density profile of KiDS~J0842+0059 is compared to the profiles of \INSPIRE\ extreme relics ($\rm DoR > 0.7$), relics ($0.34 <\rm DoR \leq 0.7$) and non-relics ($\rm DoR \leq 0.34$). Finally, in the right panel, the mass density profile of KiDS~J0842+0059 is compared to the profiles of the 3 local relics: NGC~1277, MRK~1216 and PGC~32873 \citep{Ferre-Mateu+17}.}
\label{fig:literature}
\end{figure*}

\section{Mass density profiles of relics and red nuggets}
\label{sec:profiles}

The new high-resolution AO imaging has allowed us to confirm that KiDS~J0842+0059 is an ultra-compact massive galaxy, according to the \INSPIRE\ threshold on the circularized radius ($\Re \lsim 2 \, \rm kpc$, \citealt{Tortora+18_UCMGs, Scognamiglio+20_UCMGs, Spiniello+21_INSPIRE-DR1}).

We note that we cannot determine the intrinsic ellipticity of the galaxy. As a result, the circularized radius may underestimate the true effective radius. In the most extreme case, where the galaxy is a circular inclined disk seen in projection, the actual effective radius could be as large as the major-axis value. However, given the high metallicity, $\alpha$-enhancement, and velocity dispersion of KiDS~J0842+0059, it is highly unlikely that the galaxy is disk-dominated, reducing the possible impact of such projection effect. Nevertheless, since determining the intrinsic ellipticity and effective radius of galaxies is typically challenging, we limit our analysis to projected quantities and adopt homogeneous definitions across different galaxy samples. 

The scope of this section is to compare the mass density profile of KiDS~J0842+0059 to other galaxies, at different cosmic time, which are believed to have similar formation path. We therefore only focus on the circularized radius, since for local relics and high-z red nuggets it is the one most commonly used in the literature.

We first converted the measured SB profile of KiDS~J0842+0059 into a projected mass density profile $\Sigma(R) = M/L \times I(R)$, where M/L is a constant stellar mass-to-light ratio\footnote{For simplicity, we assume no color gradients, implying a constant \ML\ as a function of radius. Even if this assumption is not entirely accurate, color gradients in massive galaxies are generally weak \citep{Tortora+11MtoLgrad}, and their impact is minimal in the K-band, which closely traces the stellar mass.} and $I(R)$ is the intensity of the best-fitting S\'ersic profile. We determined the M/L by imposing that the total mass obtained by integrating $\Sigma(R)$ to large radii is equal to the value of $10^{11}\, \rm \Msun$ we have determined from KiDS photometric data in \cite{Scognamiglio+20_UCMGs}. We then applied the same procedure to other samples. In particular, we determined the $\Sigma(R)$ for the entire \INSPIRE\ sample, including ultra-compact galaxies of any DoR, and a randomly selected sample of 500 normal-size galaxies at $z < 0.5$ with $\Re > 2.5 \, \rm kpc$ and stellar masses similar to those of \INSPIRE\ galaxies.

For these normal-sized galaxies we rely on KiDS r-band structural parameters and stellar masses obtained using KiDS photometry (\citealt{Roy+18, Tortora+20_UCMGs_env,Tortora+18_UCMGs,Scognamiglio+20_UCMGs}). Finally, we obtained the $\Sigma(R)$ profiles for the red nuggets with $\Re < 2 \, \rm kpc$ and $\mst \gsim \, 6 \times 10^{10}\, \rm \Msun$ in the sample of $z\sim 2$ quiescent galaxies in \cite{Szomoru+12}. 

We plot the distribution of the mass density profiles of these samples in the left panel of \Fig\ref{fig:literature} as shaded regions\footnote{We evaluated all the profiles at given radial values and joined all the data points. Then, we only retained points within the 16th and 84th percentile of the sample distribution in each radial bin, to exclude outlier profiles.}, and compared to KiDS~J0842+0059. KiDS~J0842+0059 (black solid line) exhibits average structural properties and surface mass density profile with respect to the \INSPIRE\ sample. What is clearly visible from this figure is that the sample of \INSPIRE\ galaxies (red shaded region) have mass density profiles which resemble perfectly those of high-$z$ red nuggets (cyan shaded region).

Selecting only relic galaxies ($\rm DoR > 0.34$, according to \citealt{Spiniello+23_INSPIRE-V}) from \INSPIRE\ does not alter these conclusions. Interestingly, when we focus on \INSPIRE\ galaxies with lower DoR values (e.g., $\rm DoR < 0.7$) or lower velocity dispersions ($< 200 \, \rm km s^{-1}$)\footnote{The most extreme relics, including KiDS~J0842+0059, exhibit higher velocity dispersions compared to UCMGs with lower DoR \citep{Spiniello+23_INSPIRE-V}.}, the scatter in the mass density profiles increases. We show this trend in the middle panel of \Fig\ref{fig:literature}, where the \INSPIRE\ sample is divided in extreme relics ($\rm DoR > 0.7$, red), relics ($0.34 <\rm DoR \leq 0.7$, green) and non-relics ($\rm DoR \leq 0.3$4, blue). The average mass density profiles do not depend on the DoR, instead their distribution is broader among galaxies with low DoR compared to the most extreme relics in the sample. 
This probably reflects the more diverse nature of these galaxies, which exhibit a broad range of star formation histories. Instead, extreme relic galaxies form a more homogeneous class of objects in terms of their stellar population properties, a consistency that also extends to their mass density profiles. In contrast, galaxies with lower DoR display more diverse properties, which likely contribute to the larger scatter observed in their stellar mass profiles. In future work, we will further investigate the stellar mass profiles of these galaxies to better understand the origin of this diversity.

Interestingly, only in the central regions ($R < 5 \, \rm kpc$) we observe a mild tendency toward shallower profiles in galaxies with lower DoR. The profiles for normal-sized galaxies at similar redshift (green shaded region in the left panel) are similar to those of ultra-compact galaxies (and in particular to those with lower DoR) at all redshifts only in the innermost region, where the compact progenitor dominates the light budget (e.g., \citealt{Barbosa21}). Instead, at larger radii ($>3-5$ kpc), compact galaxies at both $z < 0.5$ and $z \sim 2$ have smaller mass densities.

Finally, the right panel of \Fig\ref{fig:literature} compares the mass density profile of KiDS~J0842+0059 with that of the three local relics. According to the DoR definition provided by \cite{Spiniello+23_INSPIRE-V} NGC~1277 has a $\rm DoR=0.95$, therefore classified as the most extreme relic found so far. KiDS~J0842+0059 ($\rm DoR=0.73$) exhibits a mass density profile that falls between that of NGC~1277 and those of the other two local relics, MRK~1216 and PGC~32873. The latter two, with their more (relatively) extended star formation histories, as found in \cite{Ferre-Mateu+17}, have higher mass densities at all radii, suggesting they accreted more mass over cosmic time. They would have a slightly lower DoR according to the \INSPIRE\ definition, but would still be classified as relics. The most extreme relics, such as NGC~1277 and KiDS~J0842+0059, have not accreted significant ex-situ mass throughout their entire cosmic histories. The behavior of Mrk~1216 and PGC~32873, which both exhibit lower DoR than NGc~1277, is consistent with \INSPIRE\ galaxies with lower-DoR.

We note that although all profiles would be shallower when surface mass density is plotted as a function of the major-axis radius instead of the circularized one (as done here), their relative differences would change only negligibly and therefore the conclusions of this comparison remain unaffected.

Summarizing the results, in terms of mass density profile, the extreme relic J0842+0059 and NGC~1277 are the perfect descendant of high-z red nuggets. Relics with more prolonged SFH (such as MRK~1216 and PGC~32873) and normal-sized galaxies have shallower mass density profile, which is an indication of an accreted ex-situ mass component that is instead missing in low-z extreme relics.

\section{Conclusions}\label{sec:conclusions}

With the \INSPIRE\ Project we have 
assembled a catalogue of 52 confirmed \UCMGs\ at $z=0.2-0.4$. 
Thanks to X-Shooter@VLT spectra, we have obtained, for each system, a precise estimate of the integrated velocity dispersion \citep{DAgo+23_INSPIRE-III}, stellar age, metal-content, and star formation history \citep{Spiniello+20_INSPIRE, Spiniello+21_INSPIRE-DR1, Spiniello+21_INSPIRE_Messenger, Spiniello+23_INSPIRE-V}, as well as individual elemental abundance and an estimate of the low-mass end of the IMF slope \citep{Martin-Navarro+23_INSPIRE-IV, Maksymowicz-Maciata+24_INSPIRE-VI}. Out of the 52 \UCMGs, 38 have been classified as relic galaxies, having formed 75\% or more of their stellar mass at early cosmic times ($z>2$). This sample
increases by more than 3 times the number of spectroscopically confirmed relic
galaxies in the nearby 
Universe. However, the photometrical 
selection of these galaxies was based on ground-based seeing
dominated observations, which make their sizes and their shapes and morphologies possibly affected by large uncertainties. Observations with high spatial resolution are necessary to firmly confirm the ultra-compact nature of these systems, case-by-case, allowing to compare their shape and morphological properties with similar galaxies in the local Universe as well as at very high redshift, and hence draw a time evolution of the most compact massive galaxies in the Universe. 

Therefore, using Adaptive Optics (AO) Ks-band observations obtained with SOUL@LBT, we targeted one of the most extreme relics in the \INSPIRE\ sample, KiDS~J0842+0059.

In particular, for KiDS~J0842+0059 ground-based r-band observations
indicated a size of $\Re \sim 1 \, \rm kpc$ and an extremely flattened light profile with an axis ratio of
$q \sim 0.3$. These measurements, however, could have been highly affected by the fact that the FWHM of these images is much bigger than the apparent size of the object \citep{Tortora+18_UCMGs}. In this paper, using LUCI@LBT AO supported observations with PSF FWHM of $0.14$ arcsec on the target, we unambiguously confirm 
the ultra-compact nature of this galaxy: KiDS~J0842+0059 is a very compact galaxy with a circularized effective
radius of $0.214 \pm 0.001$ arcsec, corresponding to $\Re \sim 1 \, \rm
kpc$, and presents disky isophotes with $q = 0.24 \pm 0.1$. Although based on a single system, the confirmation of the compactness of this extreme relic supports the goodness of the \INSPIRE\ selection procedure and estimate of  effective radii with seeing-limited KiDS data.

The structural properties and disky isophotal shapes of KiDS~J0842+0059 fully resemble 
those accurately measured for the three local
relic galaxies in \cite{Ferre-Mateu+17}, for which HST data is available. They are also very similar to those of
high-redshift red nuggets (e.g.,
\citealt{Damjanov+09_rednuggets,Szomoru+12,Baker23,Glazebrook+24}). We have also shown that KiDS~J0842+0059 has a mass density profile which resembles that of NGC~1277 and of high-z red-nuggets. On the contrary, UCMGs with a relatively more prolongued SFH (lower DoR) present a larger scatter in their distribution, compared to the most extreme relics. Both UCMGs with lower DoR at $R \lsim 5 \rm kpc$ and normal-sized galaxies have shallower mass profiles, which might be 
explained by the formation of ex-situ stars (accreted or later formed), a kind of process which is instead completely missing in the most extreme relics in the Universe, as KiDS~J0842+0059 and NGC~1277.

Our results are thus perfectly consistent with the "two-phase" scenario, where relic galaxies have stopped forming stars and have missed the merging phase (e.g., \citealt{Oser+10}).

Combined with spectroscopic evidence showing that KiDS~J0842+0059 had assembled its entire stellar mass by $z=2$, and is characterized by metal-rich, $\alpha$-enhanced stellar populations \citep{Spiniello+21_INSPIRE-DR1} and a dwarf-rich IMF \citep{Maksymowicz-Maciata+24_INSPIRE-VI}, we have now reached an unambiguous confirmation that KiDS~J0842+0059 is a remnant of a high-$z$ red nugget.

This is the first time that high-resolution imaging is obtained
for a relic galaxy outside the local Universe
with such a high data quality and
spatial resolution. KiDS~J0842+0059 is the relic galaxy with the
most complete set of observations (seeing-dominated imaging in optical and NIR,
high-quality spectroscopy and high-resolution NIR photometry), and
thus, up to now, the most secure relic and compact galaxy beyond the
local Universe. However, obtaining structural parameter measurements with AO- or space-based data for the remaining galaxies in the \INSPIRE\ sample remains crucial to definitively confirm their compactness, ensuring the uniqueness of this relic galaxy sample. The Euclid Wide Survey \citep{Scaramella22,EuclidSkyOverview} will help us taking a first step in this direction, providing high spatial resolution data on the 17 \INSPIRE\ galaxies in the Southern Hemisphere, as well as observing more than 30\,000 new UCMGs, extending the redshift boundaries up to $z\sim2$. This will finally allow us to bridge the gap between the two phases of the massive galaxies formation scenario \citep{Oser+10}.  Furthermore, recently, a proposal to follow up the most extreme relic, the largest and smaller UCMGs and the object with the largest velocity dispersion in \INSPIRE\ has been approved. We have been awarded MUSE-IFS data in the Narrow Field Mode (NFM) AO configuration ($0.025''$/pixel with a $7.5''\times7.5''$ field-of-view) to get unprecedently precise structure parameters and to resolve the internal structure and kinematics of relics for the first time ever beyond the local Universe.

\section*{Acknowledgements}

We thank the anonymous referee for their suggestions, which helped improve the paper. CT acknowledges the INAF grant 2022 LEMON. GT acknowledges financial support from the European Research
Council (ERC) Advanced Grant under the European Union's Horizon Europe research and innovation programme (grant agreement AdG GALPHYS, No. 101055023). GD acknowledges support by UKRI-STFC grants: ST/T003081/1 and ST/X001857/1. AFM has received support from RYC2021-031099-I and PID2021-123313NA-I00 of MICIN/AEI/10.13039/501100011033/FEDER,UE, NextGenerationEU/PRT. DS carried out this research at the Jet Propulsion Laboratory, California Institute of Technology, under a contract with the National Aeronautics and Space Administration (80NM0018D0004). FLB acknowledges support from INAF minigrant 1.05.23.04.01.

%%%%%%%%%%%%%%%%%%%%%%%%%%%%%%%%%%%%%%%%%%%%%%%%%%
\section*{Data Availability}
The \INSPIRE\ catalogue is publicly available through the ESO Phase 3 
Archive Science Portal under the collection \INSPIRE\ (\url{https://archive.eso.org/scienceportal/home?data_collection=INSPIRE},\url{https:https://doi.eso.org/10.18727/archive/36}). The LUCI-SOUL@LBT data used in this paper, will be made available to the interested readers upon request to the authors.



\bibliographystyle{mnras}
\bibliography{myrefs} % if your bibtex file is called example.bib

\begin{thebibliography}{}
\makeatletter
\relax
\def\mn@urlcharsother{\let\do\@makeother \do\$\do\&\do\#\do\^\do\_\do\%\do\~}
\def\mn@doi{\begingroup\mn@urlcharsother \@ifnextchar [ {\mn@doi@} {\mn@doi@[]}}
\def\mn@doi@[#1]#2{\def\@tempa{#1}\ifx\@tempa\@empty \href {http://dx.doi.org/#2} {doi:#2}\else \href {http://dx.doi.org/#2} {#1}\fi \endgroup}
\def\mn@eprint#1#2{\mn@eprint@#1:#2::\@nil}
\def\mn@eprint@arXiv#1{\href {http://arxiv.org/abs/#1} {{\tt arXiv:#1}}}
\def\mn@eprint@dblp#1{\href {http://dblp.uni-trier.de/rec/bibtex/#1.xml} {dblp:#1}}
\def\mn@eprint@#1:#2:#3:#4\@nil{\def\@tempa {#1}\def\@tempb {#2}\def\@tempc {#3}\ifx \@tempc \@empty \let \@tempc \@tempb \let \@tempb \@tempa \fi \ifx \@tempb \@empty \def\@tempb {arXiv}\fi \@ifundefined {mn@eprint@\@tempb}{\@tempb:\@tempc}{\expandafter \expandafter \csname mn@eprint@\@tempb\endcsname \expandafter{\@tempc}}}

\bibitem[\protect\citeauthoryear{{Agapito}, {Arcidiacono}, {Quir{\'o}s-Pacheco}  \& {Esposito}}{{Agapito} et~al.}{2014}]{Agapito+14}
{Agapito} G.,  {Arcidiacono} C.,  {Quir{\'o}s-Pacheco} F.,   {Esposito} S.,  2014, \mn@doi [Experimental Astronomy] {10.1007/s10686-014-9380-7}, \href {https://ui.adsabs.harvard.edu/abs/2014ExA....37..503A} {37, 503}

\bibitem[\protect\citeauthoryear{{Baker} et~al.,}{{Baker} et~al.}{2023}]{Baker23}
{Baker} W.~M.,  et~al., 2023, \mn@doi [arXiv e-prints] {10.48550/arXiv.2306.02472}, \href {https://ui.adsabs.harvard.edu/abs/2023arXiv230602472B} {p. arXiv:2306.02472}

\bibitem[\protect\citeauthoryear{{Barbosa}, {Spiniello}, {Arnaboldi}, {Coccato}, {Hilker}  \& {Richtler}}{{Barbosa} et~al.}{2021}]{Barbosa21}
{Barbosa} C.~E.,  {Spiniello} C.,  {Arnaboldi} M.,  {Coccato} L.,  {Hilker} M.,   {Richtler} T.,  2021, \mn@doi [\aap] {10.1051/0004-6361/202039809}, \href {https://ui.adsabs.harvard.edu/abs/2021A&A...649A..93B} {649, A93}

\bibitem[\protect\citeauthoryear{{Beasley}, {Trujillo}, {Leaman}  \& {Montes}}{{Beasley} et~al.}{2018}]{Beasley18}
{Beasley} M.~A.,  {Trujillo} I.,  {Leaman} R.,   {Montes} M.,  2018, \mn@doi [\nat] {10.1038/nature25756}, \href {https://ui.adsabs.harvard.edu/abs/2018Natur.555..483B} {555, 483}

\bibitem[\protect\citeauthoryear{{Beltramo-Martin}, {Correia}, {Neichel}  \& {Fusco}}{{Beltramo-Martin} et~al.}{2018}]{Beltramo-Martin+18}
{Beltramo-Martin} O.,  {Correia} C.~M.,  {Neichel} B.,   {Fusco} T.,  2018, \mn@doi [\mnras] {10.1093/mnras/sty2399}, \href {https://ui.adsabs.harvard.edu/abs/2018MNRAS.481.2349B} {481, 2349}

\bibitem[\protect\citeauthoryear{{Britton}}{{Britton}}{2006}]{Britton06}
{Britton} M.~C.,  2006, \mn@doi [\pasp] {10.1086/505547}, \href {https://ui.adsabs.harvard.edu/abs/2006PASP..118..885B} {118, 885}

\bibitem[\protect\citeauthoryear{{Buitrago} et~al.,}{{Buitrago} et~al.}{2018}]{Buitrago+18_compacts}
{Buitrago} F.,  et~al., 2018, \mn@doi [\aap] {10.1051/0004-6361/201833785}, \href {http://adsabs.harvard.edu/abs/2018A%26A...619A.137B} {619, A137}

\bibitem[\protect\citeauthoryear{{Cantalloube} et~al.,}{{Cantalloube} et~al.}{2020}]{Cantalloube+20}
{Cantalloube} F.,  et~al., 2020, \mn@doi [\aap] {10.1051/0004-6361/201937397}, \href {https://ui.adsabs.harvard.edu/abs/2020A&A...638A..98C} {638, A98}

\bibitem[\protect\citeauthoryear{{Charbonnier} et~al.,}{{Charbonnier} et~al.}{2017}]{Charbonnier+17_compact_galaxies}
{Charbonnier} A.,  et~al., 2017, \mn@doi [\mnras] {10.1093/mnras/stx1142}, \href {http://adsabs.harvard.edu/abs/2017MNRAS.469.4523C} {469, 4523}

\bibitem[\protect\citeauthoryear{{Comer{\'o}n} et~al.,}{{Comer{\'o}n} et~al.}{2023}]{Comeron23}
{Comer{\'o}n} S.,  et~al., 2023, \mn@doi [\aap] {10.1051/0004-6361/202346291}, \href {https://ui.adsabs.harvard.edu/abs/2023A&A...675A.143C} {675, A143}

\bibitem[\protect\citeauthoryear{{D'Ago} et~al.,}{{D'Ago} et~al.}{2023}]{DAgo+23_INSPIRE-III}
{D'Ago} G.,  et~al., 2023, \mn@doi [\aap] {10.1051/0004-6361/202245542}, \href {https://ui.adsabs.harvard.edu/abs/2023A&A...672A..17D} {672, A17}

\bibitem[\protect\citeauthoryear{{Damjanov} et~al.,}{{Damjanov} et~al.}{2009}]{Damjanov+09_rednuggets}
{Damjanov} I.,  et~al., 2009, \mn@doi [\apj] {10.1088/0004-637X/695/1/101}, \href {https://ui.adsabs.harvard.edu/abs/2009ApJ...695..101D} {695, 101}

\bibitem[\protect\citeauthoryear{{Damjanov}, {Geller}, {Zahid}  \& {Hwang}}{{Damjanov} et~al.}{2015}]{Damjanov+15_compacts}
{Damjanov} I.,  {Geller} M.~J.,  {Zahid} H.~J.,   {Hwang} H.~S.,  2015, \mn@doi [\apj] {10.1088/0004-637X/806/2/158}, \href {http://adsabs.harvard.edu/abs/2015ApJ...806..158D} {806, 158}

\bibitem[\protect\citeauthoryear{{Euclid Collaboration: Mellier}, {Abdurro'uf}, {Acevedo~Barroso}  et~al.}{{Euclid Collaboration: Mellier} et~al.}{2024}]{EuclidSkyOverview}
{Euclid Collaboration: Mellier} Y.,  {Abdurro'uf} {Acevedo~Barroso} J.,   et~al., 2024, \aap, accepted, \href {https://ui.adsabs.harvard.edu/abs/2024arXiv240513491E} {p. arXiv:2405.13491}

\bibitem[\protect\citeauthoryear{{Euclid Collaboration} et~al.,}{{Euclid Collaboration} et~al.}{2022}]{Scaramella22}
{Euclid Collaboration} et~al., 2022, \mn@doi [\aap] {10.1051/0004-6361/202141938}, \href {https://ui.adsabs.harvard.edu/abs/2022A&A...662A.112E} {662, A112}

\bibitem[\protect\citeauthoryear{{Ferr{\'e}-Mateu}, {Mezcua}, {Trujillo}, {Balcells}  \& {van den Bosch}}{{Ferr{\'e}-Mateu} et~al.}{2015}]{Ferre-Mateu+15}
{Ferr{\'e}-Mateu} A.,  {Mezcua} M.,  {Trujillo} I.,  {Balcells} M.,   {van den Bosch} R.~C.~E.,  2015, \mn@doi [\apj] {10.1088/0004-637X/808/1/79}, \href {http://adsabs.harvard.edu/abs/2015ApJ...808...79F} {808, 79}

\bibitem[\protect\citeauthoryear{{Ferr{\'e}-Mateu}, {Trujillo}, {Mart{\'{\i}}n-Navarro}, {Vazdekis}, {Mezcua}, {Balcells}  \& {Dom{\'{\i}}nguez}}{{Ferr{\'e}-Mateu} et~al.}{2017}]{Ferre-Mateu+17}
{Ferr{\'e}-Mateu} A.,  {Trujillo} I.,  {Mart{\'{\i}}n-Navarro} I.,  {Vazdekis} A.,  {Mezcua} M.,  {Balcells} M.,   {Dom{\'{\i}}nguez} L.,  2017, \mn@doi [\mnras] {10.1093/mnras/stx171}, \href {http://adsabs.harvard.edu/abs/2017MNRAS.467.1929F} {467, 1929}

\bibitem[\protect\citeauthoryear{{Flores-Freitas}, {Chies-Santos}, {Furlanetto}, {De Rossi}, {Ferreira}, {Zenocratti}  \& {Alamo-Mart{\'\i}nez}}{{Flores-Freitas} et~al.}{2022}]{Flores-Freitas22}
{Flores-Freitas} R.,  {Chies-Santos} A.~L.,  {Furlanetto} C.,  {De Rossi} M.~E.,  {Ferreira} L.,  {Zenocratti} L.~J.,   {Alamo-Mart{\'\i}nez} K.~A.,  2022, \mn@doi [\mnras] {10.1093/mnras/stac187}, \href {https://ui.adsabs.harvard.edu/abs/2022MNRAS.512..245F} {512, 245}

\bibitem[\protect\citeauthoryear{{Glazebrook} et~al.,}{{Glazebrook} et~al.}{2024}]{Glazebrook+24}
{Glazebrook} K.,  et~al., 2024, \mn@doi [\nat] {10.1038/s41586-024-07191-9}, \href {https://ui.adsabs.harvard.edu/abs/2024Natur.628..277G} {628, 277}

\bibitem[\protect\citeauthoryear{{Haguenauer}, {Agapito}  \& {Neichel}}{{Haguenauer} et~al.}{2022}]{Haguenauer+22}
{Haguenauer} P.,  {Agapito} G.,   {Neichel} B.,  2022, in {Schreiber} L.,  {Schmidt} D.,   {Vernet} E.,  eds,  Society of Photo-Optical Instrumentation Engineers (SPIE) Conference Series Vol. 12185, Adaptive Optics Systems VIII. p. 1218565, \mn@doi{10.1117/12.2627243}

\bibitem[\protect\citeauthoryear{{Heidt} et~al.,}{{Heidt} et~al.}{2018}]{Heidt+18_LUCI}
{Heidt} J.,  et~al., 2018, in Ground-based and Airborne Instrumentation for Astronomy VII. p. 107020B, \mn@doi{10.1117/12.2313506}

\bibitem[\protect\citeauthoryear{{Hill}}{{Hill}}{2010}]{Hill10_LBT}
{Hill} J.~M.,  2010, \mn@doi [\ao] {10.1364/AO.49.00D115}, \href {http://adsabs.harvard.edu/abs/2010ApOpt..49..115H} {49, 115}

\bibitem[\protect\citeauthoryear{{Hopkins}, {Bundy}, {Murray}, {Quataert}, {Lauer}  \& {Ma}}{{Hopkins} et~al.}{2009}]{Hopkins+09_compacts}
{Hopkins} P.~F.,  {Bundy} K.,  {Murray} N.,  {Quataert} E.,  {Lauer} T.~R.,   {Ma} C.-P.,  2009, \mn@doi [\mnras] {10.1111/j.1365-2966.2009.15062.x}, \href {https://ui.adsabs.harvard.edu/abs/2009MNRAS.398..898H} {398, 898}

\bibitem[\protect\citeauthoryear{{La Barbera}, {de Carvalho}, {Kohl-Moreira}, {Gal}, {Soares-Santos}, {Capaccioli}, {Santos}  \& {Sant'anna}}{{La Barbera} et~al.}{2008}]{LaBarbera_08_2DPHOT}
{La Barbera} F.,  {de Carvalho} R.~R.,  {Kohl-Moreira} J.~L.,  {Gal} R.~R.,  {Soares-Santos} M.,  {Capaccioli} M.,  {Santos} R.,   {Sant'anna} N.,  2008, \mn@doi [\pasp] {10.1086/588614}, \href {http://adsabs.harvard.edu/abs/2008PASP..120..681L} {120, 681}

\bibitem[\protect\citeauthoryear{{Labb{\'e}} et~al.,}{{Labb{\'e}} et~al.}{2023}]{Labbe+23}
{Labb{\'e}} I.,  et~al., 2023, \mn@doi [\nat] {10.1038/s41586-023-05786-2}, \href {https://ui.adsabs.harvard.edu/abs/2023Natur.616..266L} {616, 266}

\bibitem[\protect\citeauthoryear{{Lisiecki}, {Ma{\l}ek}, {Siudek}, {Pollo}, {Krywult}, {Karska}  \& {Junais}}{{Lisiecki} et~al.}{2023}]{Lisiecki+23}
{Lisiecki} K.,  {Ma{\l}ek} K.,  {Siudek} M.,  {Pollo} A.,  {Krywult} J.,  {Karska} A.,   {Junais} 2023, \mn@doi [\aap] {10.1051/0004-6361/202243616}, \href {https://ui.adsabs.harvard.edu/abs/2023A&A...669A..95L} {669, A95}

\bibitem[\protect\citeauthoryear{{Maksymowicz-Maciata} et~al.,}{{Maksymowicz-Maciata} et~al.}{2024}]{Maksymowicz-Maciata+24_INSPIRE-VI}
{Maksymowicz-Maciata} M.,  et~al., 2024, \mn@doi [\mnras] {10.1093/mnras/stae1318}, \href {https://ui.adsabs.harvard.edu/abs/2024MNRAS.531.2864M} {531, 2864}

\bibitem[\protect\citeauthoryear{{Mart{\'\i}n-Navarro}, {van de Ven}  \& {Y{\i}ld{\i}r{\i}m}}{{Mart{\'\i}n-Navarro} et~al.}{2019}]{Martin-Navarro19}
{Mart{\'\i}n-Navarro} I.,  {van de Ven} G.,   {Y{\i}ld{\i}r{\i}m} A.,  2019, \mn@doi [\mnras] {10.1093/mnras/stz1544}, \href {https://ui.adsabs.harvard.edu/abs/2019MNRAS.487.4939M} {487, 4939}

\bibitem[\protect\citeauthoryear{{Mart{\'\i}n-Navarro} et~al.,}{{Mart{\'\i}n-Navarro} et~al.}{2023a}]{Martn-Navarro+23_ISPIRE-IV}
{Mart{\'\i}n-Navarro} I.,  et~al., 2023a, \mn@doi [\mnras] {10.1093/mnras/stad503}, \href {https://ui.adsabs.harvard.edu/abs/2023MNRAS.521.1408M} {521, 1408}

\bibitem[\protect\citeauthoryear{{Mart{\'\i}n-Navarro} et~al.,}{{Mart{\'\i}n-Navarro} et~al.}{2023b}]{Martin-Navarro+23_INSPIRE-IV}
{Mart{\'\i}n-Navarro} I.,  et~al., 2023b, \mn@doi [\mnras] {10.1093/mnras/stad503}, \href {https://ui.adsabs.harvard.edu/abs/2023MNRAS.521.1408M} {521, 1408}

\bibitem[\protect\citeauthoryear{{Moura}, {Chies-Santos}, {Furlanetto}, {Zhu}  \& {Canossa-Gosteinski}}{{Moura} et~al.}{2024}]{Moura+24}
{Moura} M.~T.,  {Chies-Santos} A.~L.,  {Furlanetto} C.,  {Zhu} L.,   {Canossa-Gosteinski} M.~A.,  2024, \mn@doi [\mnras] {10.1093/mnras/stae013}, \href {https://ui.adsabs.harvard.edu/abs/2024MNRAS.528..353M} {528, 353}

\bibitem[\protect\citeauthoryear{{Naab}, {Johansson}  \& {Ostriker}}{{Naab} et~al.}{2009}]{Naab+09}
{Naab} T.,  {Johansson} P.~H.,   {Ostriker} J.~P.,  2009, \mn@doi [\apjl] {10.1088/0004-637X/699/2/L178}, \href {http://adsabs.harvard.edu/abs/2009ApJ...699L.178N} {699, L178}

\bibitem[\protect\citeauthoryear{{Neichel} et~al.,}{{Neichel} et~al.}{2021}]{Neichel+21_TIPTOP}
{Neichel} B.,  et~al., 2021, in Adaptive Optics Systems VII. p. 114482T (\mn@eprint {arXiv} {2101.06486}), \mn@doi{10.1117/12.2561533}

\bibitem[\protect\citeauthoryear{{Oser}, {Ostriker}, {Naab}, {Johansson}  \& {Burkert}}{{Oser} et~al.}{2010}]{Oser+10}
{Oser} L.,  {Ostriker} J.~P.,  {Naab} T.,  {Johansson} P.~H.,   {Burkert} A.,  2010, \mn@doi [\apj] {10.1088/0004-637X/725/2/2312}, \href {http://adsabs.harvard.edu/abs/2010ApJ...725.2312O} {725, 2312}

\bibitem[\protect\citeauthoryear{Pinna et~al.,}{Pinna et~al.}{2019}]{Pinna+2019_SOUL}
Pinna E.,  et~al., 2019, in {Adaptive Optics for Extremely Large Telescopes 6 - Conference Proceedings}. \url {http://ao4elt6.copl.ulaval.ca/proceedings/401-yw8a-251.pdf}

\bibitem[\protect\citeauthoryear{{Pinna} et~al.,}{{Pinna} et~al.}{2023}]{Pinna+23_SOUL}
{Pinna} E.,  et~al., 2023, in Adaptive Optics for Extremely Large Telescopes (AO4ELT7). p.~80 (\mn@eprint {arXiv} {2310.14447}), \mn@doi{10.13009/AO4ELT7-2023-082}

\bibitem[\protect\citeauthoryear{{Quilis} \& {Trujillo}}{{Quilis} \& {Trujillo}}{2013}]{Quilis_Trujillo13}
{Quilis} V.,  {Trujillo} I.,  2013, \mn@doi [\apjl] {10.1088/2041-8205/773/1/L8}, \href {http://adsabs.harvard.edu/abs/2013ApJ...773L...8Q} {773, L8}

\bibitem[\protect\citeauthoryear{{Roy} et~al.,}{{Roy} et~al.}{2018}]{Roy+18}
{Roy} N.,  et~al., 2018, \mn@doi [\mnras] {10.1093/mnras/sty1917}, \href {http://adsabs.harvard.edu/abs/2018MNRAS.480.1057R} {480, 1057}

\bibitem[\protect\citeauthoryear{{Salvador-Rusi{\~n}ol}, {Ferr{\'e}-Mateu}, {Vazdekis}  \& {Beasley}}{{Salvador-Rusi{\~n}ol} et~al.}{2022}]{Salvador-Rusinol22}
{Salvador-Rusi{\~n}ol} N.,  {Ferr{\'e}-Mateu} A.,  {Vazdekis} A.,   {Beasley} M.~A.,  2022, \mn@doi [\mnras] {10.1093/mnras/stac2070}, \href {https://ui.adsabs.harvard.edu/abs/2022MNRAS.515.4514S} {515, 4514}

\bibitem[\protect\citeauthoryear{{Scognamiglio} et~al.,}{{Scognamiglio} et~al.}{2020}]{Scognamiglio+20_UCMGs}
{Scognamiglio} D.,  et~al., 2020, \mn@doi [\apj] {10.3847/1538-4357/ab7db3}, \href {https://ui.adsabs.harvard.edu/abs/2020ApJ...893....4S} {893, 4}

\bibitem[\protect\citeauthoryear{{Scognamiglio} et~al.,}{{Scognamiglio} et~al.}{2024}]{Scognamiglio+24_INSPIREVII}
{Scognamiglio} D.,  et~al., 2024, \mn@doi [\mnras] {10.1093/mnras/stae2185}, \href {https://ui.adsabs.harvard.edu/abs/2024MNRAS.534.1597S} {534, 1597}

\bibitem[\protect\citeauthoryear{{Spiniello}, {Tortora}, {D'Ago}, {Napolitano}  \& {Inspire Team}}{{Spiniello} et~al.}{2021a}]{Spiniello+21_INSPIRE_Messenger}
{Spiniello} C.,  {Tortora} C.,  {D'Ago} G.,  {Napolitano} N.~R.,   {Inspire Team} 2021a, \mn@doi [The Messenger] {10.18727/0722-6691/5241}, \href {https://ui.adsabs.harvard.edu/abs/2021Msngr.184...26S} {184, 26}

\bibitem[\protect\citeauthoryear{{Spiniello} et~al.,}{{Spiniello} et~al.}{2021b}]{Spiniello+20_INSPIRE}
{Spiniello} C.,  et~al., 2021b, \mn@doi [\aap] {10.1051/0004-6361/202038936}, \href {https://ui.adsabs.harvard.edu/abs/2021A&A...646A..28S} {646, A28}

\bibitem[\protect\citeauthoryear{{Spiniello} et~al.,}{{Spiniello} et~al.}{2021c}]{Spiniello+21_INSPIRE-DR1}
{Spiniello} C.,  et~al., 2021c, \mn@doi [\aap] {10.1051/0004-6361/202140856}, \href {https://ui.adsabs.harvard.edu/abs/2021A&A...654A.136S} {654, A136}

\bibitem[\protect\citeauthoryear{{Spiniello} et~al.,}{{Spiniello} et~al.}{2024}]{Spiniello+23_INSPIRE-V}
{Spiniello} C.,  et~al., 2024, \mn@doi [\mnras] {10.1093/mnras/stad3703}, \href {https://ui.adsabs.harvard.edu/abs/2024MNRAS.527.8793S} {527, 8793}

\bibitem[\protect\citeauthoryear{{Stone}, {Arora}, {Courteau}  \& {Cuillandre}}{{Stone} et~al.}{2021}]{Stone+2021_Autoprof}
{Stone} C.~J.,  {Arora} N.,  {Courteau} S.,   {Cuillandre} J.-C.,  2021, \mn@doi [\mnras] {10.1093/mnras/stab2709}, \href {https://ui.adsabs.harvard.edu/abs/2021MNRAS.508.1870S} {508, 1870}

\bibitem[\protect\citeauthoryear{{Szomoru}, {Franx}  \& {van Dokkum}}{{Szomoru} et~al.}{2012}]{Szomoru+12}
{Szomoru} D.,  {Franx} M.,   {van Dokkum} P.~G.,  2012, \mn@doi [\apj] {10.1088/0004-637X/749/2/121}, \href {http://adsabs.harvard.edu/abs/2012ApJ...749..121S} {749, 121}

\bibitem[\protect\citeauthoryear{{Tortora}, {Napolitano}, {Romanowsky}, {Jetzer}, {Cardone}  \& {Capaccioli}}{{Tortora} et~al.}{2011}]{Tortora+11MtoLgrad}
{Tortora} C.,  {Napolitano} N.~R.,  {Romanowsky} A.~J.,  {Jetzer} P.,  {Cardone} V.~F.,   {Capaccioli} M.,  2011, \mn@doi [\mnras] {10.1111/j.1365-2966.2011.19438.x}, \href {http://adsabs.harvard.edu/abs/2011MNRAS.418.1557T} {418, 1557}

\bibitem[\protect\citeauthoryear{{Tortora} et~al.,}{{Tortora} et~al.}{2016}]{Tortora+16_compacts_KiDS}
{Tortora} C.,  et~al., 2016, \mn@doi [\mnras] {10.1093/mnras/stw184}, \href {http://adsabs.harvard.edu/abs/2016MNRAS.457.2845T} {457, 2845}

\bibitem[\protect\citeauthoryear{{Tortora} et~al.,}{{Tortora} et~al.}{2018}]{Tortora+18_UCMGs}
{Tortora} C.,  et~al., 2018, \mn@doi [\mnras] {10.1093/mnras/sty2564}, \href {http://adsabs.harvard.edu/abs/2018MNRAS.481.4728T} {481, 4728}

\bibitem[\protect\citeauthoryear{{Tortora} et~al.,}{{Tortora} et~al.}{2020}]{Tortora+20_UCMGs_env}
{Tortora} C.,  et~al., 2020, \mn@doi [\aap] {10.1051/0004-6361/202038373}, \href {https://ui.adsabs.harvard.edu/abs/2020A&A...638L..11T} {638, L11}

\bibitem[\protect\citeauthoryear{{Trujillo}, {Cenarro}, {de Lorenzo-C{\'a}ceres}, {Vazdekis}, {de la Rosa}  \& {Cava}}{{Trujillo} et~al.}{2009}]{Trujillo+09_superdense}
{Trujillo} I.,  {Cenarro} A.~J.,  {de Lorenzo-C{\'a}ceres} A.,  {Vazdekis} A.,  {de la Rosa} I.~G.,   {Cava} A.,  2009, \mn@doi [\apjl] {10.1088/0004-637X/692/2/L118}, \href {http://adsabs.harvard.edu/abs/2009ApJ...692L.118T} {692, L118}

\bibitem[\protect\citeauthoryear{{Trujillo}, {Ferr{\'e}-Mateu}, {Balcells}, {Vazdekis}  \& {S{\'a}nchez-Bl{\'a}zquez}}{{Trujillo} et~al.}{2014}]{Trujillo+14}
{Trujillo} I.,  {Ferr{\'e}-Mateu} A.,  {Balcells} M.,  {Vazdekis} A.,   {S{\'a}nchez-Bl{\'a}zquez} P.,  2014, \mn@doi [\apjl] {10.1088/2041-8205/780/2/L20}, \href {http://adsabs.harvard.edu/abs/2014ApJ...780L..20T} {780, L20}

\bibitem[\protect\citeauthoryear{{Turchi}, {Masciadri}  \& {Veillet}}{{Turchi} et~al.}{2022}]{Turchi+22}
{Turchi} A.,  {Masciadri} E.,   {Veillet} C.,  2022, in {Marshall} H.~K.,  {Spyromilio} J.,   {Usuda} T.,  eds,  Society of Photo-Optical Instrumentation Engineers (SPIE) Conference Series Vol. 12182, Ground-based and Airborne Telescopes IX. p. 121824O, \mn@doi{10.1117/12.2629813}

\bibitem[\protect\citeauthoryear{{Wellons} et~al.,}{{Wellons} et~al.}{2016}]{Wellons16}
{Wellons} S.,  et~al., 2016, \mn@doi [\mnras] {10.1093/mnras/stv2738}, \href {https://ui.adsabs.harvard.edu/abs/2016MNRAS.456.1030W} {456, 1030}

\bibitem[\protect\citeauthoryear{{Y{\i}ld{\i}r{\i}m}, {van den Bosch}, {van de Ven}, {Mart{\'\i}n-Navarro}, {Walsh}, {Husemann}, {G{\"u}ltekin}  \& {Gebhardt}}{{Y{\i}ld{\i}r{\i}m} et~al.}{2017}]{Yildrim17}
{Y{\i}ld{\i}r{\i}m} A.,  {van den Bosch} R. C.~E.,  {van de Ven} G.,  {Mart{\'\i}n-Navarro} I.,  {Walsh} J.~L.,  {Husemann} B.,  {G{\"u}ltekin} K.,   {Gebhardt} K.,  2017, \mn@doi [\mnras] {10.1093/mnras/stx732}, \href {https://ui.adsabs.harvard.edu/abs/2017MNRAS.468.4216Y} {468, 4216}

\bibitem[\protect\citeauthoryear{{Zolotov} et~al.,}{{Zolotov} et~al.}{2015}]{Zolotov+15_nuggets}
{Zolotov} A.,  et~al., 2015, \mn@doi [\mnras] {10.1093/mnras/stv740}, \href {https://ui.adsabs.harvard.edu/abs/2015MNRAS.450.2327Z} {450, 2327}

\bibitem[\protect\citeauthoryear{{de Jong} et~al.,}{{de Jong} et~al.}{2015}]{deJong+15_KiDS_paperI}
{de Jong} J.~T.~A.,  et~al., 2015, \mn@doi [\aap] {10.1051/0004-6361/201526601}, \href {http://adsabs.harvard.edu/abs/2015A%26A...582A..62D} {582, A62}

\bibitem[\protect\citeauthoryear{{de Jong} et~al.,}{{de Jong} et~al.}{2017}]{deJong+17_KiDS_DR3}
{de Jong} J.~T.~A.,  et~al., 2017, \mn@doi [\aap] {10.1051/0004-6361/201730747}, \href {http://adsabs.harvard.edu/abs/2017A%26A...604A.134D} {604, A134}

\makeatother
\end{thebibliography}

%%%%%%%%%%%%%%%%%%%%%%%%%%%%%%%%%%%%%%%%%%%%%%%%%%

%%%%%%%%%%%%%%%%% APPENDICES %%%%%%%%%%%%%%%%%%%%%

\appendix

\section{PSF determination}\label{app:PSF_full}

\begin{table}
\centering
\caption{PSF model parameters.}\label{tab:PSFparams}
\begin{tabular}{llc}
\hline
Type & Parameter & Value\\
\hline
\multirow{8}{*}{Fixed} & Telescope Elevation [deg] & 57$\pm$1 \\
{} & DIMM seeing [arcsec] & 0.9$\pm$0.1 \\
{} & Outer Scale [m] & 25 \\
{} & Ground Wind Speed [m/s] & 4$\pm$1 \\
{} & Guide Star Magnitude (R+I) & 16 \\
{} & Pupil Sampling [pixel] & 13$\times$13 \\
{} & Loop Frame Rate [Hz] & 358 \\
{} & Loop Gain (tilt) & 0.6 \\
{} & Loop Gain (high order modes) & 0.25 \\
\hline
\multirow{3}{*}{Fitting} & Seeing at zenith [arcsec] & 0.72 \\
{} & Isoplanatic Angle [arcsec] & 2.01 \\
{} & Average Wind Speed [m/s] & 19 \\
\hline
\end{tabular}
\end{table}

\begin{figure*}
    \begin{center}
%    \begin{tabular}{c}
\includegraphics[width=0.51\textwidth]{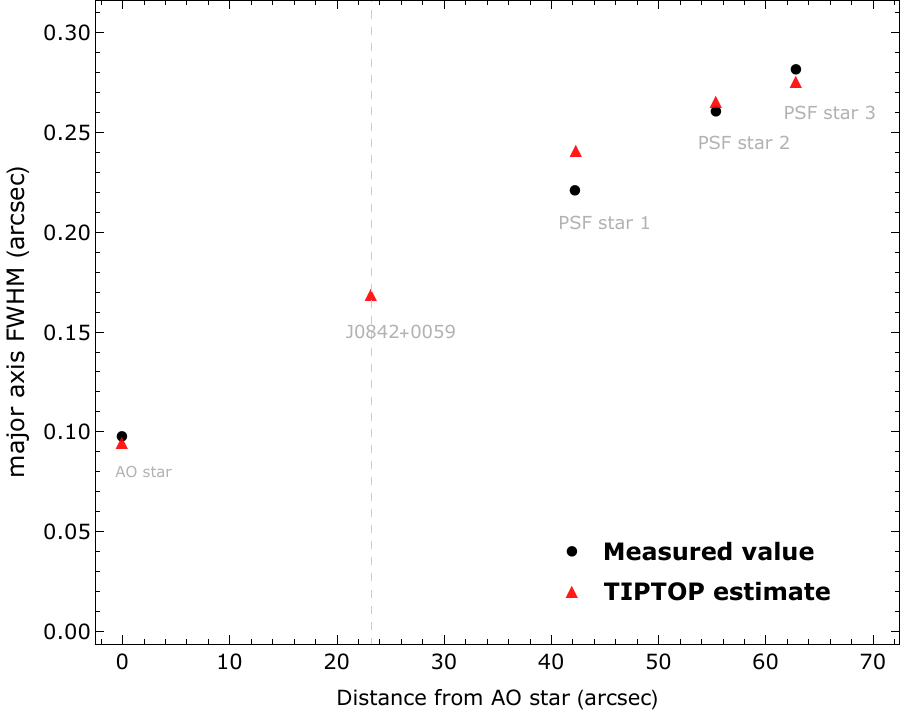}
\includegraphics[width=0.44\textwidth]{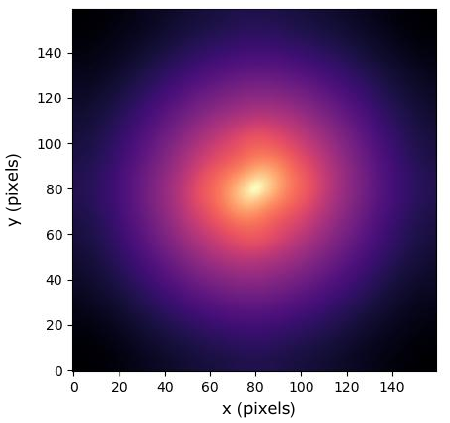}
%    \end{tabular}
    \end{center}
	 \caption{\label{fig:FWHM_est} {\it Left}. FWHM of the PSFs at the position of the 4 stars and the galaxy for the major axis only. Comparison between the values estimated from LUCI images (black dots) and the TIPTOP estimate (red triangles). Only the TIPTOP estimate is available at 23.2 arcsec, where the galaxy is located (dashed line). {\it Right.} Image of the PSF, used in the structural parameter derivation procedure.}
\end{figure*}

We selected \textit{TIPTOP} (\citealt{Neichel+21_TIPTOP}) as simulation software. Our approach consists of two steps. First, we fitted the 4 stars (AO and PSF stars) with TIPTOP, and then we extrapolated the PSF to the galaxy position. 
The PSF modelling includes mainly four atmospheric parameters: a) seeing, b) outer scale, c) vertical turbulence profile and d) wind profile, and the telescope and AO loop parameters. As merit function for the fit, we used the uncertainty 
on the FWHM of the 4 stars. 
The complete set of TIPTOP parameters is summarised in Tab.~\ref{tab:PSFparams}, which lists in the first block the parameters we kept fixed (first block) and those we fit (second block), with the assumptions we will describe below.

The 3 PSF stars in the field have a distinct elongation at $\sim$45 degrees: this is expected given their relative position to the AO guide star. 
The AO guide star PSF also has a small amount of elongation, about a factor of 1.2, along 
the 45 degree direction. This is probably due to the tip/tilt oscillation caused by some vibrations in the telescope path. Moreover,  there is a triangular shape near the core of the AO guide star PSF, which is caused by a non-common path aberration between SOUL and LUCI. 
We neglected such features in the PSF determination because they have a small effect on the off-axis PSF.

We used the AO-star FWHM to estimate the TIPTOP parameters associated with the magnitude of the oscillations, corresponding to a 2D Gaussian kernel with a FWHM of 0.068 and 0.056 arcsec on the major and minor axis, respectively. Note that these values are compatible but slightly smaller than the residual tip/tilt jitter estimated by the AO telemetry (0.03 arcsec RMS corresponds to a FWHM of 0.07 arcsec). They allow us to obtain the correct FWHM of the AO guide star ($\sim 0.10$ and $\sim 0.08$ arcsec for major and minor axis respectively) in the range of seeing values considered for the fit ($\sim 0.9$ arcsec). 
In fact, the AO guide star FWHM depends mainly on the seeing value and on the vibrations level, it is marginally affected by the outer scale, the wind speed and direction (the wind profile affects the halo of the PSF as can be seen in \citealt{Cantalloube+20}) and it is not affected by the vertical turbulence (C$_n^2$) profile, which, instead, strongly affects off-axis stars/objects.

We decided to simplify the estimation of the atmospheric profiles chosing 
a 4-layer configuration with heights 103, 725, 2637 and 11068 m as in \cite{Agapito+14} and we changed the vertical turbulence (C$_n^2$) profile to produce different isoplanatic angles. The wind speed profile was chosen using as the ground wind speed the value measured by the telescope anemometers, 4 m/s. For the wind speed in the higher layers, we explored a range of values close to those produced by the ALTA forecasts \citep{Turchi+22}: the jet stream speed during the night was predicted to be between 40 and 60 m/s. 
Finally, we decided to fix the outer scale at a value of 25 m, which is compatible with the statistics of this parameter estimated by the Adaptive Optics Facility at the ESO VLT between 2019 and 2022 \citep{Haguenauer+22}.

The results in terms of FWHM of the major axis of the PSFs are summarized in \Fig\ref{fig:FWHM_est}. The first set of PSFs is taken from the LUCI images where field stars are present, while the second set of PSFs is the one generated by TIPTOP at the location of the field star and of the galaxy. The FWHM values are computed using a fit to a Moffat function. Note that the major axis is aligned with the lines connecting the AO star and the field stars.

We decided to not use the FWHM of the minor axis of the field stars in this fit. 
This is because it has a lower sensitivity to the distance from the AO star than the major axis, and a number of bad pixels prevent us from getting an accurate estimate of this FWHM on one of the three field stars. However, this decision is not significant for the result, as the FWHM of the minor axis has a strong correlation with the major axis and therefore does not provide independent elements to the fit \citep{Britton06,Beltramo-Martin+18}.

The PSF image is shown in the right panel of Figure~\ref{fig:FWHM_est}. This is clearly stretched in the upper-right direction in the SOUL tile, which is in the direction of the AO star.

Finally, we note that the estimation of the PSF does not take into account the temporal variations of the telescope elevation, the atmospheric parameters (seeing, outer scale, vertical turbulence and wind profile) and the structural vibrations, since we have considered average values.

\begin{figure*}
\centering
\includegraphics[width=0.99\textwidth]{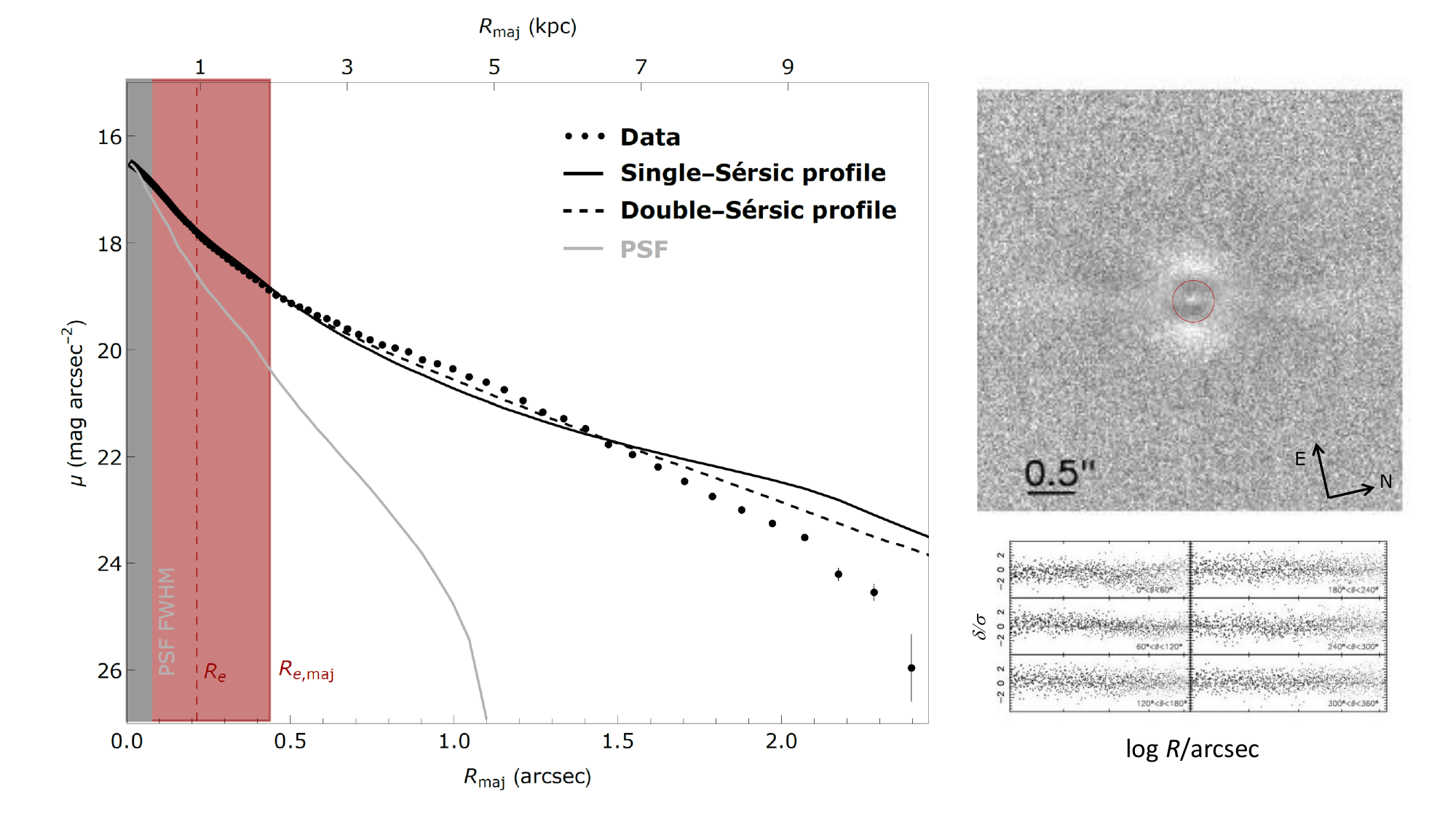}
\caption{{\it Left}. The same as \Fig\ref{fig:SBprofiles}, but with in addition the best-fitting PSF-convolved model obtained by fitting a double-S\'ersic profile using \twodphot, plotted as dashed black line. {\it Right}. Residual image generated by \twodphot\ for the double-S\'ersic fit. The red circle marks the circularized \Re\ of the single-S\'ersic model. See \Fig\ref{fig:SBprofiles} for more details.}\label{fig:SBprofiles_appendix} 
\end{figure*}

\section{Double-S\'ercic fit}\label{app:double_sersic}

We have also fitted to the data a combination of a S\'ersic and an exponential profile. The best-fitting model and the residual are shown in \Fig\ref{fig:SBprofiles_appendix}.

The fit gives $n = 3.82$,
$q = 0.58$ and $R_{e,circ} = 0.05$ arcsec for the innermost component and
$q = 0.138$ and $R_{e,circ} = 0.31$ arcsec for the second one, for which n is fixed to 1. This fit returns a $\chi^{2} = 1.2$, slightly lower than the value obtained for the single-S\'ersic fit. For simplicity, we refer to the more concentrated component as a bulge, and we call the $n=1$ component a disk. This allows us to define a bulge-to-total parameter, as the ratio of the total flux in the bulge and total flux in the galaxy. The bulge component is dominating over the disk with a $B/T = 0.68$. Accordingly to the smaller $\chi^2$, the
residuals shown in the right panel of \Fig \ref{fig:SBprofiles_appendix} exhibit a slight improvement compared to the single-S\'ersic case. However, since the central component primarily fits the inner SB profile, where the PSF has the greatest influence, we consider this solution less meaningful and tend to exclude it.

\bsp	% typesetting comment
\label{lastpage}
\end{document}